\documentclass[10pt,twocolumn,letterpaper]{article}

\usepackage{cvpr}
\usepackage{times}
\usepackage{epsfig}
\usepackage{graphicx}
\usepackage{amsmath}
\usepackage{amssymb}
\usepackage{graphicx}
\usepackage{float}
\usepackage{subfigure}
\usepackage{multirow}
\usepackage{siunitx}
\usepackage{bm}
\usepackage{authblk}
\usepackage{cite}

\usepackage[breaklinks=true,bookmarks=false]{hyperref}

\cvprfinalcopy 

\def\cvprPaperID{****} 

\begin{document}

\def\andname{,}
\author[1]{Jing Zhou}
\author[2]{Akira Nakagawa}
\author[2]{Keizo Kato}
\author[1]{Sihan Wen}
\author[2]{Kimihiko Kazui}
\author[1]{Zhiming Tan}
{
	\makeatletter
	\renewcommand\AB@affilsepx{, \protect\Affilfont}
	\makeatother	
	\affil[1]{Fujitsu R\&D Center Co. Ltd.}
	\affil[2]{Fujitsu Laboratories Ltd.}
}
{
	\makeatletter
	\let\AB@affilsep\AB@affilsepx
	\makeatother	
	\affil[1]{\small \{zhoujing, wensihan, zhmtan\}@cn.fujitsu.com,}
}
\affil[2]{\small \{anaka, kato.keizo, kazui.kimihiko\}@fujitsu.com}
\title{Variable Rate Image Compression Method with Dead-zone Quantizer}
\renewcommand\Authands{, }

\maketitle
\pagestyle{empty}

\begin{abstract}
Deep learning based image compression methods have achieved superior performance compared with transform based conventional codec. With end-to-end Rate-Distortion Optimization (RDO) in the codec, compression model is optimized with Lagrange multiplier $\lambda$. For conventional codec, signal is decorrelated with orthonormal transformation, and uniform quantizer is introduced. We propose a variable rate image compression method with dead-zone quantizer. Firstly, the autoencoder network is trained with RaDOGAGA \cite{radogaga} framework, which can make the latents isometric to the metric space, such as SSIM and MSE. Then the conventional dead-zone quantization method with arbitrary step size is used in the common trained network to provide the flexible rate control. With dead-zone quantizer, the experimental results show that our method performs comparably with independently optimized models within a wide range of bitrate.   
\end{abstract}

\section{Introduction}
Image compression is a kind of traditional and well-studied technique. With the key challenge of rate and distortion tradeoff, traditional codecs, such as JPEG \cite{jpeg} and JPEG2000 \cite{jp2k}, usually break the pipeline into 3 modules: transformation, quantization, and entropy codec. Joint optimization over rate and distortion has longly been considered as an intractable problem. For transform coding \cite{transform}, components are optimized separately, then they are fit together manually. Taking JPEG for example, it uses 8x8 block in Discrete Cosine Transform (DCT), then adapts run-length encoding to exploit the sparsity pattern of extracted frequency coefficients. Quantization is applied on coefficients to realize different compression level.

With the development of deep learning technology, more and more methods have been proposed to realize end-to-end rate distortion tradeoff \cite{balle2017end, balle2018variational, Fabian, 	minnen2018joint, joint, hyper, lee2018context}. Autoencoder has been proved to reduce dimensionality effectively by introducing an $'information$ $bottleneck'$ that forces the network to find and exploit redundancies. Variational AutoEncoder (VAE) was evolved to model the underlying data distribution explicitly with sacrifice of distortion.
To some extent the autoencoder-based compression methods \cite{balle2018variational, theis2017lossy} can be seen as VAEs, while the entropy model for rate estimation corresponds to the prior on the latents.

In these RDO methods, the Lagrange multiplier $\lambda$ is introduced to modulate tradeoff between rate and distortion as $R+\lambda \cdot D$. A common way for rate adaptation is training multiple models with different $\lambda$, which is very tedious. To deal with such limitation, many methods have been proposed \cite{RNN, theis2017lossy, samsung, modulated}. Toderici et al. \cite{RNN} proposed a RNN based progressive encoding and decoding scheme to generate target quality from low to high level, which requires large hardware storage and high performance in application.
Choi et al. \cite{samsung} combines the Lagrange multiplier and quantization bin size to realize rate control. But the range of quantizer is constrained within a small range to reduce performance degradation. To cover a broad rate range, multiple models trained with different $\lambda$ are used. 

Taking traditional codec into consideration, if neural network based image compression can be trained to realize orthonormal transformation, such as Karhunen-Lo{\'e}ve transform (KLT) and DCT, we can introduce the conventional quantization method to the neural network based orthonormal encoder. Inspired from this, we propose a variable rate image compression method with dead-zone quantizer \cite{dead-zone}. RaDOGAGA \cite{radogaga} can realize an orthonormal latent space with minimal entropy, such as KLT and DCT.
 Because of orthonormality (orthogonal and uniform scaling) and minimal entropy, the common network can be used for an arbitrary quantizer. Firstly, the autoencoder network is trained with RaDOGAGA with metrics such as Mean Square Error (MSE), and Multi-Scale Structural SIMilarity (MS-SSIM), et al. Latents isometric to these metric spaces are derived. In inference, a uniform quantizer is utilized to obtain variable rates with different step sizes as traditional codec with fixed encoder/decoder. To further improve its performance, the dead-zone quantizer is introduced. According to our experimental results, our method can achieve comparable R-D performance with models optimized separately both in PSNR and MS-SSIM.

\begin{figure}[t]
     	\centering
		\includegraphics[width=1\linewidth]{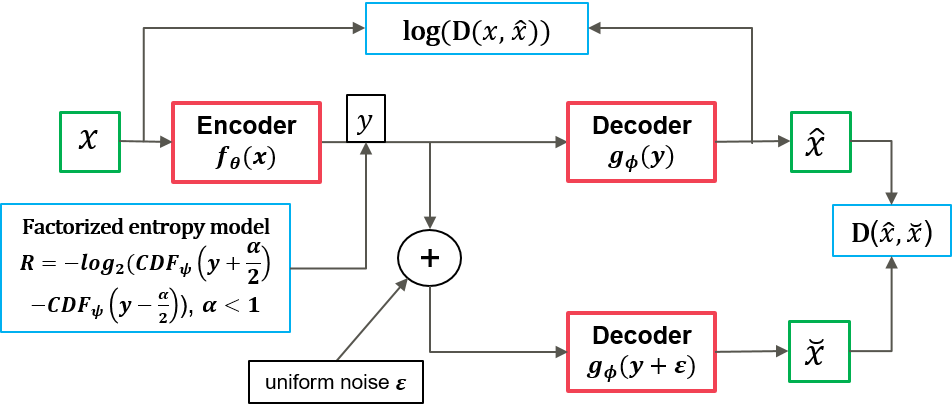}
	\caption{The RaDOGAGA framework for training}
	\label{fig:1}
\end{figure} 

\section{Variable rate image compression}

\subsection{Training framework: RaDOGAGA} 
As for RaDOGAGA, which is a rate-distortion optimization guided autoencoder, it proves that deep autoencoder can achieve orthonormal transform with RDO. Just like common compression framework, it mainly contains encoder $f_{\theta}(\bm x)$, decoder $g_{\phi}(y)$ and factorized entropy model Cumulative Density Function $CDF_{\psi}(y)$ with parameters $\theta$, $\phi$ and $\psi$. For the factorized entropy model as described in \cite{balle2018variational}, bounds of latents $y$ should be finite and known ahead. The bounds for latents are also trained, which contains not only the maximum and minimum ranges, but also median value as a kind of quantization offset. The probability can be calculated as follows from $CDF_{\psi}(y)$, where $\alpha \textless 1$.
\begin{equation}
   P_{\psi}(y) = CDF_{\psi}(y+ \frac{\alpha}{2}) - CDF_{\psi}(y-\frac{\alpha}{2})
   \label{Prob}
\end{equation} 

 Compared with the mainstream $R+\lambda \cdot D$ loss, the difference can be illustrated from two aspects. The first is that two inputs exist for the decoder $g_{\phi}(y)$. One is the extracted latents $y$, and the corresponding output is $\hat {\bm x}$. The other is noised latents $y+\epsilon$, where $\epsilon$ is uniform noise with variance ${\sigma}$, and the corresponding output is $\breve{\bm x}$. The second aspect is loss function shown in Eq. \ref{loss}.
\begin{equation}
L = -log_2(P_{\psi}(y)) + \lambda_{1} \cdot h(D(\bm x,\hat{\bm x})) + \lambda_{2} \cdot D(\hat{\bm x}, \breve{\bm x})
\label{loss}
\end{equation}

The first distortion $D(\bm x,\hat{\bm x})$ is for reconstruction, and the second $D(\hat{\bm x}, \breve{\bm x})$ aims to influence the scaling, i.e., Jacobian matrix. According to Rol{\'i}neck et al. \cite{Rolinek}, $ D(x,\breve{\bm x}) \simeq D(\bm x,\hat{\bm x}) + D(\hat{\bm x}, \breve{\bm x})$. Here we use $h(d)=log(d)$, which can encourage better reconstruction and  orthogonality. $D(\cdot)$ is an arbitrary metric, such as MSE, SSIM, or Binary Cross Entropy (BCE), etc. 

As shown in RaDOGAGA, the latent becomes orthonormal to the metric defined image space with minimal entropy such as KLT. Let $D({\bm x,\hat {\bm x}})$ be a metric such as SSIM, MSE, and so on.	
Usually $D({\bm x,{\bm x}+{\delta {\bm x}}})$ can be approximated by ${}^T {\delta {\bm x}} \bm A_{\bm x} {\delta {\bm x}}$ where ${\delta {\bm x}}$ is an arbitrary micro displacement and $\bm A_{\bm x}$ is a Riemannian metric tensor at $\bm x$. By using RaDOGAGA, each row vector of Jacobi matrix becomes orthonormal at any $\bm x$ in the inner product space with metric $\bm A_{\bm x}$ as follows, where $\delta_{ij}$ denotes Kronecker's delta.	

\begin{equation}
{}^T (\frac{\partial{\bm x}} {\partial{y_i}}) \bm A_{\bm x} (\frac{\partial{\bm x}} {\partial{y_j}}) = \frac{1}{2 \lambda_2 {\sigma}^2} \delta_{ij}
\end{equation}

\subsection{Compression framework with quantization}
In this section, we'll explain a compression framework with an arbitrary step sized quantization by using the common encoder $f_{\theta}(x)$, decoder $g_{\phi}(y)$ and entropy model $CDF_{\psi} (y,Q)$. After training with RaDOGAGA, the latents are optimized to be orthonormal to metric defined inner product space.
For example, if trained with MS-SSIM, the space of latent is isometric against Riemannian manifold whose metric is MS-SSIM. Thus rate control can be realized as conventional way. 

\begin{figure}[htbp]
	\centering
	\subfigure[The encoder procedure]{
	   \includegraphics[width=1\linewidth]{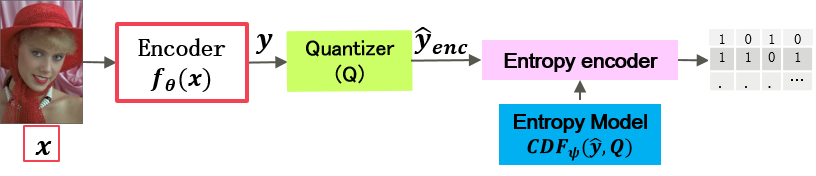}
       \label{fig:enc}}       
    \quad
    \subfigure[The decoder procedure]{
       \includegraphics[width=1\linewidth]{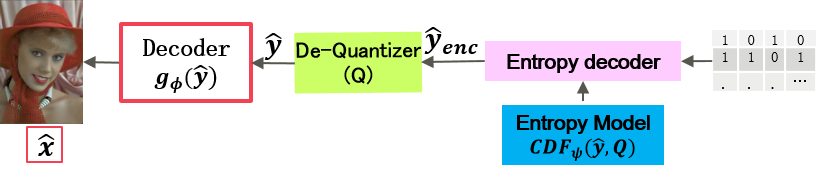}
       \label{fig:dec}}
    \caption{The compression framework}
    \label{fig:test}
\end{figure}

Firstly, input signal is decorrelated by orthonormal transformation with the $Encoder$. Then decorrelated data $y$ can be quantized uniformly with arbitrary Q, short for quantization step size. Then quantized symbols $\hat{y}_{enc}$ are entropy coded using the $Entropy$ $Model$ after $Quantizer$. In the decoder procedure, the symbols are entropy decoded losslessly using the arithmetic coding with the estimated probability, dequantized, and fed to the $Decoder$ as shown in Figure \ref{fig:dec}. Thus, with arbitrary Q, flexible rate control can be realized easily.

\subsection{Dead-zone quantizer}
\begin{figure}[H]
	\centering
	\includegraphics[width=1\linewidth]{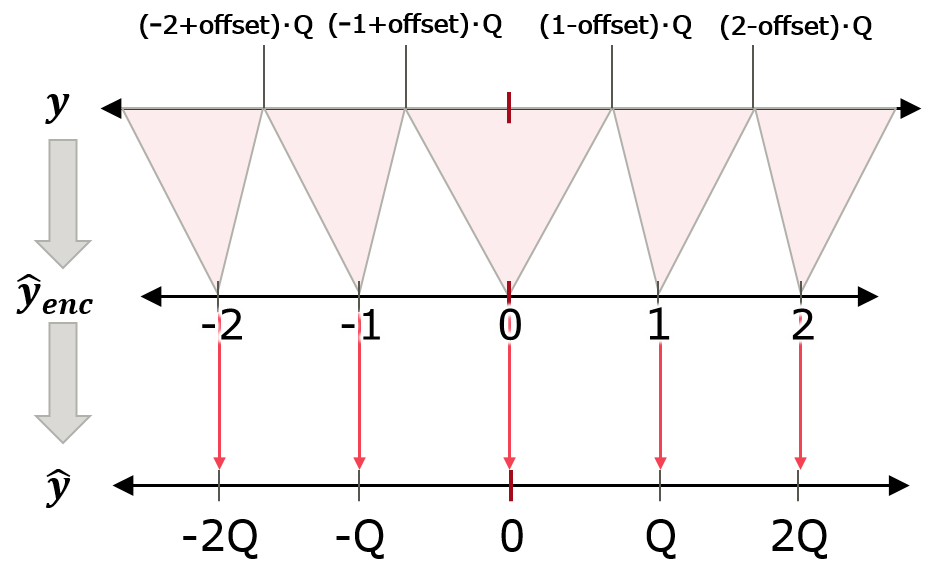}
	\caption{The dead zone quantizer.}
	\label{fig:dead-q}
\end{figure}

We employ a simple dead-zone quantizer widely used in image compression such as H.264 as shown in Figure \ref{fig:dead-q}. Based on the trained $CDF_{\psi}$, quantizer, and dead-zone offset, the probability of each latent in $\hat{y}$ is estimated in advance for each representative.

Compressed symbols $\hat{y}_{enc}$ for entropy codec is shown in Eq. \ref{y_enc} with dead-zone quantizer, where $sgn(\cdot)$ is the signum function. $y$ is centered on the median value from trained $CDF_{\psi}$, while $0$ is used for simplicity. The $offset$ can be from 0 to 0.5. If $offset=0.5$, it means round quantization.
\begin{equation}
\hat{y}_{enc} = sgn(y) \cdot \lfloor \frac{\left\vert y \right\vert}{Q} + offset \rfloor
\label{y_enc}
\end{equation}

Then $\hat{y}$ can be obtained with $De$-$Quantizer(Q)$.
\begin{equation}
\hat{y} = \hat{y}_{enc} \cdot Q
\label{q}
\end{equation}
Figure \ref{fig:dead-q} shows the upper $y_{upper}$ and lower $y_{lower}$ bound of $y$ for each quantized representative value. A probability for a quantize symbol $\hat{y}$ is estimated with equations as follow, where $0<\omega<1$. 
\begin{equation}
y_{upper} = (\hat{y}_{enc} + 0.5 + sgn(\hat{y}_{enc}+\omega) \times(0.5 - offset)) \cdot Q
\label{upper}
\end{equation}
\begin{equation}
y_{lower} = (\hat{y}_{enc} - 0.5 + sgn(\hat{y}_{enc}-\omega) \times (0.5 - offset)) \cdot Q
\label{lower}
\end{equation} 
\begin{equation}
P_{\hat{y}_{enc}} = CDF(y_{upper}) - CDF(y_{lower})
\label{prob-y}
\end{equation}

\section{Experiments}
\subsection{Training method}
The training dataset contains more than 6000 images. Some of them are from lossless dataset, such as dataset from Workshop and Challenge on Learned Image Compression (CLIC)\footnote[1]{https://www.compression.cc} and DIV2K, a super-resolution dataset. Others are from flickr.com, where images are not in lossless format with $2\times 2$ downsampling to reduce compression artefacts. In training, the network is fed with $256 \times 256\times 3$ patches cropped from these full resolution images randomly with a minibatch size of 8. Each image is normalized by dividing by 255 for each RGB channel.

To compare the performance of our variable rate compression method with independently optimized method, we use the same network structure in \cite{balle2017end}, and models are trained with its official open source code\footnote[2]{https://github.com/tensorflow/compression}. We set the bottleneck number 128. If optimized for MSE, it's calculated after scaling the normalized images to 255.  

\subsection{Conventional model with independent training}
For neural network based conventional method, such as\cite{balle2017end}, the R-D curve is pointed through models trained with different $\lambda$. Each point represents a compression level optimized with a specific R-D tradeoff $\lambda$ independently. Such method can achieve ideal performance, but it requires more memory to store and more computing to train them. In other words, the training time is multiplied. To compare the RD curves both in PSNR and MS-SSIM, different $\lambda$ and loss are used. For PSNR, MSE is distortion in training, and $\lambda \in \left\{0.001, 0.003, 0.005, 0.01, 0.02, 0.03, 0.1\right\}$. For MS-SSIM, it's the same training loss with $\lambda \in \left\{4, 8, 16, 32, 64, 96\right\}$.

\begin{figure}[htbp]
	\centering
	\subfigure[R-D curve of PSNR]{
		\includegraphics[width=1\linewidth]{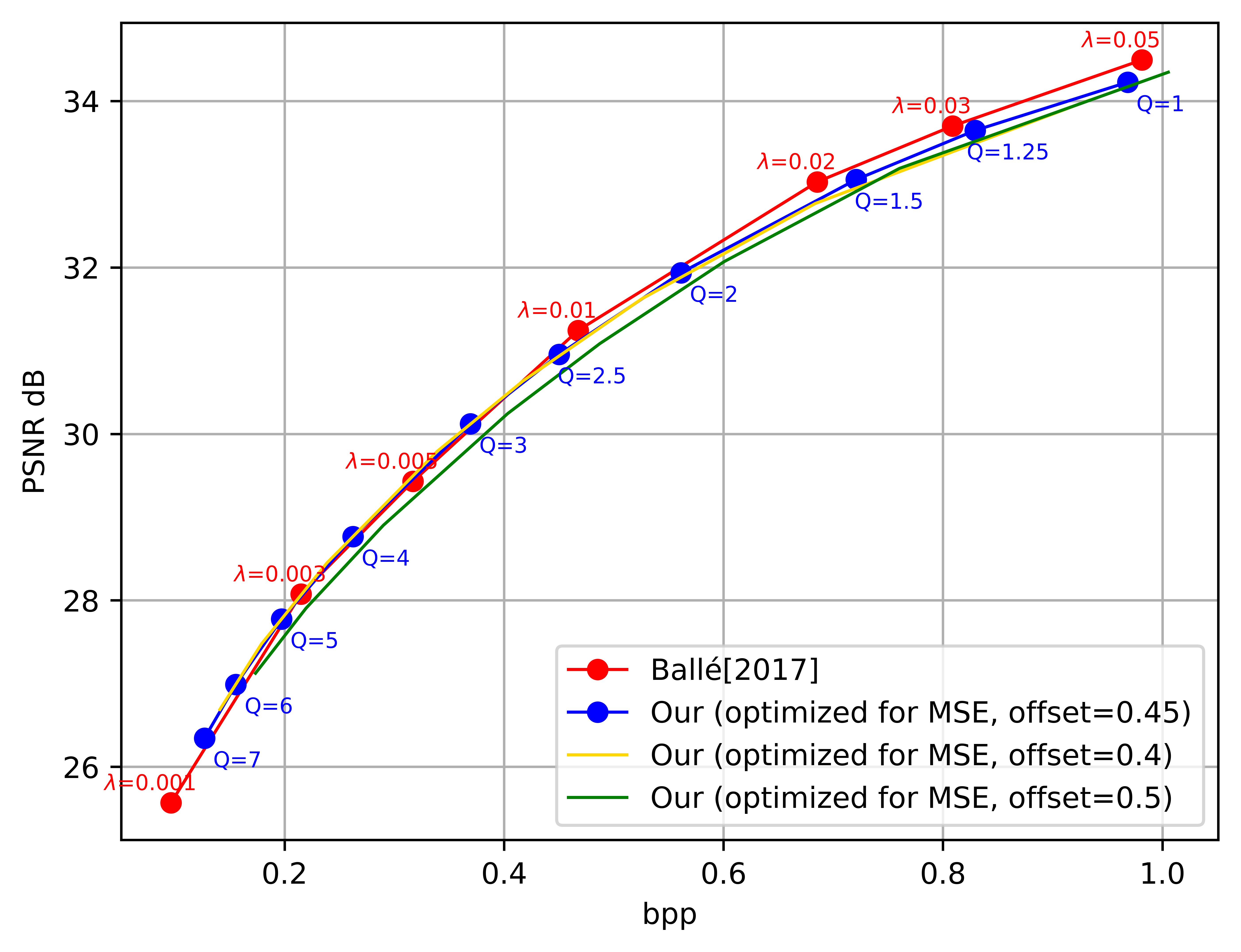}
		\label{fig:psnr}}       
	\quad
	\subfigure[R-D curve of MS-SSIM dB]{
		\includegraphics[width=1\linewidth]{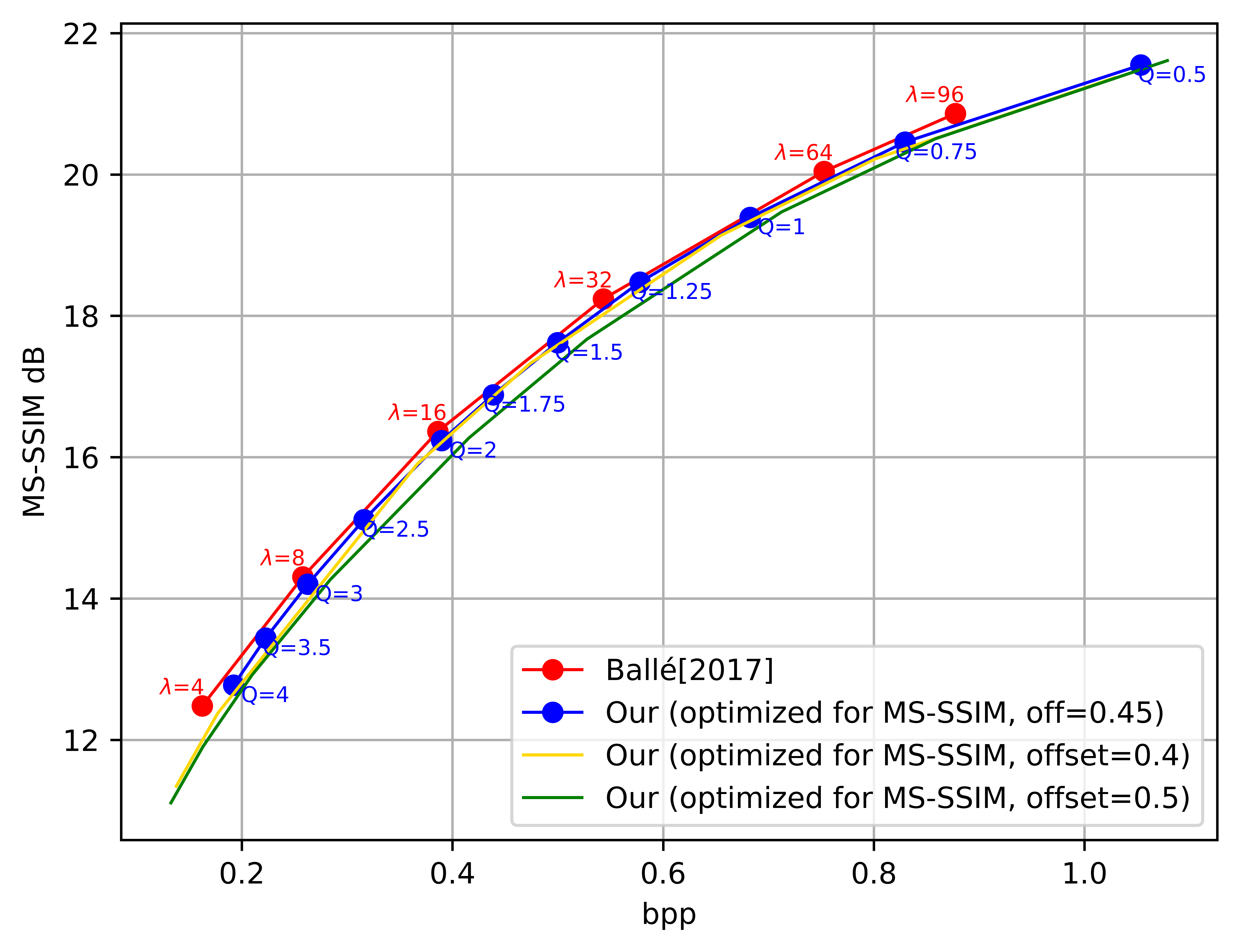}
		\label{fig:msssim}}
	\caption{PSNR \ref{fig:psnr} and MS-SSIM \ref{fig:msssim} comparison on Kodak dataset. The '$\lambda=16$' around the red curve indicates model is optimized with $R+16 \cdot D$. The blue text '$Q=1$' means that quantization step is 1.}
	\label{fig:rd}
\end{figure}

\begin{figure*}[htbp]
	\begin{center}
		\includegraphics[width=1\linewidth]{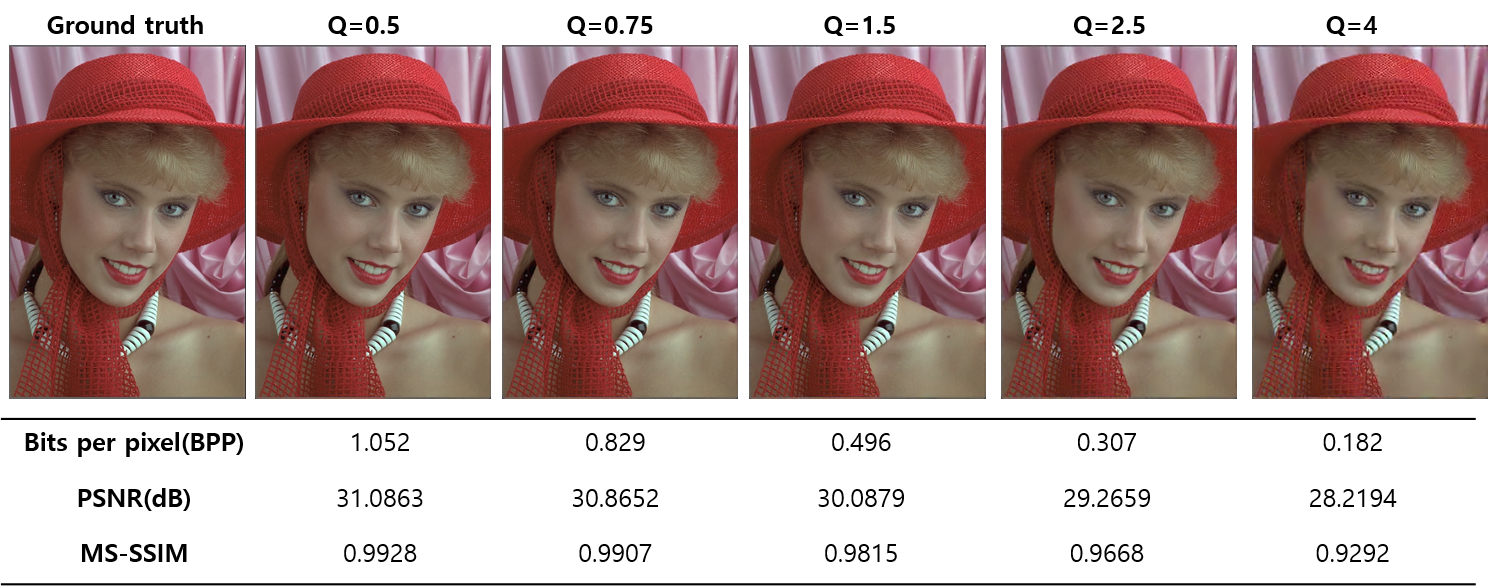}
	\end{center}
	\caption{Reconstructed images with different Q, model optimized with MS-SSIM and $offset=0.45$}
	\label{fig:kodim_04}
\end{figure*}

\subsection{Proposed model with arbitrary quantization}
We train our model as RaDOGAGA framework explained in section 2.1 with MS-SSIM and MSE. Parameters are shown in Table \ref{tab-1}. If optimized for MSE, it means two distortions in Eq. \ref{loss} are MSE in training. Once the training is completed, $Q$ in a relative wide range can be used to tune the desired rate/distortion. What's more, we use the dead-zone quantizer in compression. Different quantization step sizes are applied to obtain different rates. For example, $Q \in \left\{0.5, 0.75, 1, 1.25, 1.5, 1.75, 2, 2.5, 3, 3.5, 4\right\}$ are used in models optimized for MS-SSIM.
\begin{table}[H]
	\footnotesize
	\begin{center}
		\begin{tabular}{|c|c|c|c|}
			\hline
			Distortion & $\lambda_{1}$ & $\lambda_{2}$ & $\alpha$\\
			\hline
			MSE & 5 & 0.2 & 0.2 \\
			\hline
			MS-SSIM & 1 & 256 & 0.2 \\
			\hline
		\end{tabular}
	\end{center}
	\caption{Parameters for our model}
	\label{tab-1}
\end{table}

\subsection{Performance}
Evaluated on 24 Kodak images\footnote[3]{http://r0k.us/graphics/kodak}, the comparison results are shown in Figure \ref{fig:rd}. MS-SSIM scores are measured in dB: ${MS{\text -}SSIM}_{dB} = -10\cdot log_{10}(1 - {MS{\text -}SSIM})$. To check the influence of different offsets in dead-zone quantizer, $offset \in \left\{ 0.4, 0.45, 0.5\right\}$ are used. As we can see, our performance is comparative with independently optimized model both on PSNR and MS-SSIM with $offset=0.45$. With small offset, quantization error increases for the centered value, which will result in aggravating distortion. So appropriate offset is beneficial to improve performance.

Figure \ref{fig:kodim_04} shows the reconstructed images through our variable rate image compression method to assess their visual quality. For $Q=4$, we can find that our method can maintain the texture well, such as areas of girl's eyelash and hair. However, the independently optimized model smooths small texture to realize better R-D tradeoff. If using larger quantization step, although the objective performance does not degrade, there is mosquito noise on the crisp edge of objects, which is a common phenomenon in classical codec compressed with DCT.

\section{Conclusion}
In this paper, we propose a variable rate image compression method based on a common autoencoder. We find that RaDOGAGA framework can realize the orthonormal latent space, which can be seen as the way of getting DCT coefficients. From this point, we introduce conventional quantizer, such as the one in JPEG/MPEG. According to our experimental results, our method can vary rate without objective metric degradation. We can deploy the trained model within a broad range of rate flexibly, which will bring great convenience in practical application. In this paper, we used simple one layer model. However, the concept of isometric latent space is generic, and will be widely applicable to more complex multi-layer compression model.

{\small
\bibliographystyle{ieee_fullname}
\bibliography{egbib}
}

\end{document}